\documentclass[12pt]{JHEP3}
\usepackage{amsfonts,amsmath,youngtab}
\usepackage{graphicx}

\title{A note on dual giant gravitons in $\textrm{AdS}_{4}\times \mathbb{CP}^{3}$}
\author{Alex Hamilton\footnote{hamilton@nassp.uct.ac.za}, Jeff Murugan\footnote{jeff@nassp.uct.ac.za} and Andrea Prinsloo\footnote{andrea.prinsloo@gmail.com}\\
Cosmology and Gravity Group, \\
Department of Mathematics and Applied Mathematics, \\
University of Cape Town, \\
Private Bag, Rondebosch, 7700, \\
South Africa.}
\author{Migael Strydom\footnote{kesseki@gmail.com}\\
Department of Physics,\\
University of Cape Town,\\
Private Bag, Rondebosch, 7700,\\
South Africa.}

\abstract{We study some of the properties of dual giant gravitons - D2-branes wrapped on an $S^{2}\subset \mathrm{AdS}_{4}$ - in type IIA string theory on 
$\mathrm{AdS}_{4}\times \mathbb{CP}^{3}$. In particular we confirm that the spectrum of small fluctuations about the giant is both real and independent of the size of the graviton. We also extend previously developed techniques for attaching open strings to giants to this D2-brane giant and focus on two particular limits of the resulting string sigma model: In the pp-wave limit we quantize the string and compute the spectrum of bosonic excitations while in the semiclassical limit, we read off the fast string Polyakov action and comment on the comparison to the Landau-Lifshitz action for the dual open spin chain.}

 \preprint{UCT-CGG-081201}

\begin{document}

%-----------------------------------------------------------------------------------
\section{Introduction}
%-----------------------------------------------------------------------------------
It is hard to overstate the role that D-branes have played in our unravelling of string theory \cite{Polchinski1}. From shedding light on black hole microstates, to resolving singularities to no less than an understanding of the emergent nature of spacetime itself, these essentially string-theoretic degrees of freedom have proven invaluable. Among this, admittedly large, class of objects one subset has stood out,  at least in the context of the AdS/CFT correspondence \cite{Maldacena1}. Giant gravitons in $\mathrm{AdS}_{5} \times S^{5}$ are D3$-$branes that wrap a contractible $3-$cycle (in this case, an $S^{3}$) either in the $5-$sphere or AdS part of the geometry and whose tensive collapse is balanced by an electric or magnetic coupling to the RR $5-$form flux \cite{Giant-refs}. As $\tfrac{1}{2}-$BPS objects, both the sphere and AdS giant gravitons are protected against strong coupling corrections and consequently form an excellent set of states with which to probe the non-perturbative sector of the type IIB string theory. On the gauge theory side, a substantial and compelling body of evidence has emerged in support of the conjecture \cite{Schur} that giant gravitons are dual to specific combinations - so-called Schur polynomials - of Higgs fields in the scalar sector of the $\mathcal{N}=4$ super-Yang-Mills theory\footnote{For an excellent review of both the literature as well as the identification of giants with Schur polynomials see section 2 of \cite{deMelloKoch1}.}. Each such Schur operator has conformal dimension $\Delta$ equal to $\mathcal{R}-$charge $n$ and is labelled by a Young diagram with $n$ boxes. This association with Young diagrams allows for a considerable simplification in an understanding of the giant graviton dynamics and interactions. For an impressive collection of results ranging from a realization of Gauss' law to gravitational radiation in the Yang-Mills theory see \cite{Giant-results}. For an interesting interpretation of some of these gauge theoretic results in a cosmological context see \cite{Hamilton-Murugan1}. Despite these extraordinary developments, in the absence of a constructive proof of the AdS/CFT conjecture, a detailed understanding of the correspondence between D$-$branes and gauge theory states remains elusive.\\

\noindent
Explicit realizations of the AdS/CFT duality have generically related gauge theories to type IIB string theory on $AdS_{5} \times X$. However, recently, there has been a mini-revolution in the field, extending the duality not just to IIA string theory, but to its lift to M-theory. Building on the seminal construction of the worldvolume action for two coincident M2-branes in \cite{BLG}, Aharony, Bergman, Jafferis and Maldacena proposed an extension to multiple M2-branes \cite{ABJM}.  This so-called ABJM model consists of a superconformal $U(N)\times U(N)$ Chern-Simons theory of levels $k$ and $-k$, coupled to two pairs of chiral superfields $(A_{a},B_{\dot{a}})$ that transform in the bifundamental representation of the gauge groups.
Moreover, the theory also exhibits a natural $1/N$ expansion with $\lambda=N/k$, the ratio of the rank of the gauge group to the Chern-Simons level number, playing the role of the 't Hooft coupling. 
When $\lambda \gg 1$ the ABJM theory becomes strongly coupled and, depending on the specific values of $k$ and $N$, is best described by a dual gravitational theory that is either M-theory or type IIA string theory. Specifically, when $k\ll N^{1/5}$ the gravitational dual is M-theory on $AdS_{4}\times S^{7}/\mathbb{Z}_{k}$ and when $N^{1/5}\ll k \ll N$, it is type IIA string theory on $AdS_{4} \times \mathbb{CP}^{3}$.\\

\noindent
With over ten years hindsight as a guide to asking the correct questions, understanding of this $AdS_{4} \times \mathbb{CP}^{3}$/superconformal Chern-Simons correspondence has progressed in leaps and bounds. In particular, detailed studies of semiclassical strings, integrability of the string sigma model, the near flat and pp-wave limits as well as the giant magnon sector of the background have all been successfully carried out (see for example \cite{Semiclassical,Integrability,Near-flat,pp-wave,Giant-Magnon} and references therein). In a sense, most of the focus thus far has been on the closed string sector of the theory with  overall results agreeing remarkably well with what was known already in the $AdS_{5} \times S^{5}$ case\footnote{Although, it is worth pointing out that integrability in this $AdS_{4}/CFT_{3}$ duality appears to be more subtle than in $AdS_{5}/CFT_{4}$. We thank Chethan Krishnan for bringing the issues raised in \cite{One-loop-mismatch} to our attention.}. On the other hand, the open string sector - the study of D-branes and their dynamics - has proven to be rather nontrivial in the limited cases studied to date \cite{ABJM,Takayanagi-rings,Berenstein-membrane}.\\

\noindent
In this note, we pursue this line of study further. Specifically, we continue the analysis of the {\it dual giant graviton} in $AdS_{4} \times \mathbb{CP}^{3}$ initiated in \cite{Takayanagi-rings}. After a short discussion of the geometry, we construct the D2-brane giant graviton in section 3, and study its spectrum of small fluctuations in section 4. In section \ref{Open strings}, we attach an open string to the dual giant and compute the bosonic spectrum of the string in the pp-wave limit. We also write down the semiclassical Polyakov action for this string in a ``fast string" limit. It is this action that must be compared to the Landau-Lifshitz action of the associated open spin chain in the dual Super Chern-Simons theory. We conclude with some comments on the gauge theory operators dual to the D2-brane giant and some speculations on future directions.  

%-----------------------------------------------------------------------------------
\section{The $AdS_{4} \times \mathbb{CP}^{3}$ background}
\label{Background}
%-----------------------------------------------------------------------------------
The $\mathcal{N}=6$, three-dimensional superconformal Chern-Simons theory of ABJM is dual to M-theory on the orbifold $AdS_{4}\times S^{7}/\mathbb{Z}_{k}$. The reduction to type IIA string theory on the ten-dimensional $AdS_{4} \times \mathbb{CP}^{3}$ is implemented by first writing the $S^{7}$ as a Hopf fibration over $\mathbb{CP}^{3}$. In terms of the complex coordinates $z_{i}\in \mathbb{C}^{4}\hookleftarrow S^{7}$, the action of the $\mathbb{Z}_{k}$ acts as $z_{i}\rightarrow e^{2\pi i/k}z_{i}$. In other words, by orbifolding by $\mathbb{Z}_{k}$ the size of the $U(1)$ fibre is effectively reduced by a factor of $k$. Taking $k\rightarrow\infty$ compactifies the M-theory circle and produces the requisite geometry. From the point of view of the gauge theory, this is a 't Hooft limit where $N,k\rightarrow\infty$ with $\lambda = N/k$ fixed. In the notation of \cite{pp-wave}, the parameters of the string background and the field theory are related through $R^{2} = \pi \sqrt{2\lambda}$ and the type IIA supermultiplet consists of the metric
\begin{eqnarray}
   \nonumber   R^{-2} ds^{2} &=
   & ds^2_{AdS_4} + 4 ds^2_{\mathbb{CP}^3} \\
  &=& -\left(1+r^{2}\right)dt^{2} + \frac{dr^{2}}{(1+r^{2})} 
   + r^{2}\left(d\theta^{2}
   + \sin^{2}{\theta}d\varphi^{2}\right) \\
    && + 4d\zeta^{2} + \cos^{2}{\zeta}\left(d\theta_{1}^{2} 
   + \sin^{2}{\theta_{1}}d\varphi_{1}^{2}\right)
   + \sin^{2}{\zeta}\left(d\theta_{2}^{2} + 
   \sin^{2}{\theta_{2}}d\varphi_{2}^{2}\right)  \nonumber \\
   && + 4\cos^{2}{\zeta}\sin^{2}{\zeta}\left(d\psi 
   + \tfrac{1}{2}\cos{\theta_{1}}d\varphi_{1}
   - \tfrac{1}{2}\cos{\theta_{2}}d\varphi_{2}\right)^{2}, \nonumber
\end{eqnarray}
a dilaton 
$e^{2\Phi} = \frac{4R^{2}}{k^{2}}$, and a 4-form field strength
\begin{equation}
    F^{(4)} = -\tfrac{3}{2}kR^{2}r^{2}\sin{\theta} ~ dt \wedge dr \wedge d\theta 
    \wedge d\varphi,
\end{equation}
associated to the RR 3-form potential
\begin{equation*}
    C^{(3)} = \tfrac{1}{2}kR^{2}r^{3}\sin{\theta} ~ dt \wedge d\theta \wedge d\varphi.
\end{equation*}
There is also an RR 2-form field strength proportional to the K\"ahler form, $J$
\begin{eqnarray}
	F^{(2)} &=& (2k)  J\\
		&=& -k \left( \sin 2\zeta d\zeta \wedge (d\psi + \tfrac{1}{2} \cos \theta_1 d\varphi_1 - \tfrac{1}{2} \cos \theta_2 d\varphi_2 ) \right. \nonumber \\
	&& \left. + \tfrac{1}{2} \cos^2 \zeta \sin \theta_1 d\theta_1 \wedge d\varphi_1 + \tfrac{1}{2} \sin^2 \zeta \sin 		\theta_2 d\theta_2 \wedge d\varphi_2 \right)  \nonumber 
\end{eqnarray}
with potential
\begin{eqnarray}
	C^{(1)} = \frac{k}{2} \left[ \cos 2\zeta (d\psi + \tfrac{1}{2} \cos \theta_1 d\varphi_1 - \tfrac{1}{2} \cos \theta_2 d\varphi_2 )
	 + \tfrac{1}{2} \cos \theta_1 d\varphi_1 + \tfrac{1}{2} \cos \theta_2 d\varphi_2 \right].
\end{eqnarray}
This background preserves the 24 supersymmetries expected of a gravity dual of the 3-dimensional $\mathcal{N}=6$ gauge theory. 
%-----------------------------------------------------------------------------------
\section{Dual giant gravitons}
\label{Giant construction}
%-----------------------------------------------------------------------------------
Our construction of the dual giant configuration here follows closely that of \cite{Takayanagi-rings} so we will be deliberately cursory, pointing out only those parts of the computation relevant for the rest of our analysis. A dual giant graviton in $AdS_{4} \times \mathbb{CP}^{3}$ is a D2-brane wrapping an $S^{2}\subset AdS_{4}$. In terms of the global AdS coordinates, the D2 is extended in $t$, $\theta$ and $\varphi$ so that choosing the
static gauge, in which $\sigma^{0} \equiv \tau = t$, $\sigma^{1} = \theta$ and $\sigma^{2} = \varphi$, the D2-brane action reads
\begin{equation}
	S_{D2} = -T_2 \int d^3\sigma \ e^{-\Phi}  \sqrt{-\det ( g + 2\pi F)}  \ + \ T_2 \int \ C^{(3)} \ + \ 2\pi T_2 \int \ C^{(1)} \wedge F
\end{equation}
with tension $T_{2} = \tfrac{1}{(2\pi)^{2}}$ in units of $\alpha' = 1$.
Here $g$ is, of course, the determinant of the pullback of the 10-dimensional spacetime metric to the D2-brane worldvolume, and $F$ is the worldvolume gauge field. In order to simplify the dual giant graviton ansatz, it is convenient to make a coordinate change 
\begin{equation}
   \chi \equiv \psi + \tfrac{1}{2}\varphi_{1} - \tfrac{1}{2}\varphi_{2}, 
   ~~~~~~ \phi_{1} \equiv \varphi_{1} ~~~~ \textrm{and}
   ~~~~ \phi_{2} \equiv \varphi_{2}.
\end{equation}
We take as our ansatz $r=r_{0}$, $\zeta = \tfrac{\pi}{4}$ and $\theta_{1}=\theta_{2}=0$, with vanishing worldvolume electric field, $F=0$. Motion of the D-brane is confined entirely to the $S^1$ parameterized by $\chi$ on $\mathbb{CP}^{3}$.  The D2-brane action reduces to
\begin{equation}
  S = -a\int{dt\left\{r_{0}^{2}\sqrt{1+r_{0}^{2}-\dot{\chi}^{2}} - r_{0}^{3}\right\}},    
  ~~~~~~ \textrm{with}
  ~~ a \equiv 2\pi T_{2}kR^{2} = \frac{kR^{2}}{2\pi}.
\end{equation}
The conserved momentum $P_{\chi}$ conjugate to $\chi$ satisfies
\begin{equation} \label{dual-momentum}
p \equiv \frac{P_{\chi}}{a} = \frac{r_{0}^{2}\dot{\chi}}{\sqrt{1+r_0^{2}-\dot{\chi}^{2}}},
\end{equation}
from which it follows that $\chi = \omega_{0}t$ is a solution to the equations of motion. The Hamiltonian $H = P_{\chi}\dot{\chi} - L$ is now easily computed as
\begin{equation}
    H = a\left\{\sqrt{1+r_{0}^{2}}\sqrt{r_{0}^{4}+p^{2}} - r_{0}^{3}\right\}.
\end{equation}
As with the dual D3-brane configuration in $AdS_{5} \times S^{5}$, this energy functional has two minima, one at $r_{0} = 0$ corresponding to a point graviton\footnote{Care must be taken in setting $r_0=0$, since (\ref{dual-momentum}) is singular when $r_0=0$ and $\omega_0^2=1$.  However, it can be shown \cite{Giant-refs} that this is a sensible limit.} and the second where $r_{0}^{2} = p^{2}$.
These solutions both satisfy the BPS bound $E=P_\chi$.  It is the latter that we will call the dual giant graviton configuration in $AdS_{4} \times \mathbb{CP}^{3}$.  Note that any geodesic on $\mathbb{CP}^3$ corresponds to such a dual giant graviton, all of which lie in the $(p,0,p)$ representation of $SU(4)$.\\

\noindent
These D2-brane solutions are dual to gauge theory operators made up of a large number of scalar fields.  Specifically, they are dual to symmetric polynomials of operators with no baryonic charge, e.g. $(A_iB_j)$ \cite{Schur, ABJM}.  Indeed, such operators fall into the $(p,0,p)$ representation of the $\mathcal{R}$-symmetry group, $SU(4)$, matching the spectrum and multiplicity of the gravity states.

%-------------------------------------------------------------------------------------
\section{The spectrum of fluctuations about the D2-brane}
\label{Fluctuations}
%-------------------------------------------------------------------------------------
Having written down the classical brane configuration, we will now analyze its stability. This information is encoded in the spectrum of small fluctuations about the dual giant \cite{Fluctuations, Pirrone}.  However, it is crucial to note\footnote{We thank Kostas Skenderis for pointing this out.} that fluctuations of the worldvolume gauge field must also be taken into account\footnote{For another example where this phenomenon arises see e.g. \cite{Pirrone}.}.  

\noindent
In three dimensions, analysis of the gauge field simplifies considerably because of its duality with a massless scalar field.  To see this, note that in the limit of small worldvolume flux, the relevant part of the action is
\begin{equation}
	S_A = - (2\pi)^2 \ T_2 \int \frac{e^{-\Phi}}{2} dA \wedge *dA \ + \  2\pi T_2 \int C^{(1)} \wedge dA
\end{equation}
where $A$ is the gauge field and $F=dA$ the associated field strength.  To dualize to the scalar field, we must first change the degree of freedom to $F_{ab}$, while at the same time adding a Lagrange multiplier, $\phi$, in order to enforce the Bianchi identity
\begin{equation}
	S_F = - (2\pi)^2 \ T_2 \int e^{-\Phi}\left\{ \tfrac{1}{2} F \wedge *F + \phi \ dF \right\} \ + \  2\pi T_2 \int C^{(1)} \wedge F.
\end{equation}
Integrating out $F = -*(d\phi + \tfrac{1}{2\pi} e^{\Phi} C^{(1)})$, the dualized action is
\begin{equation}
	S_\phi = - (2\pi)^2 \ T_2 \int d^3\sigma \sqrt{-g} \  e^{-\Phi} \left\{ \frac{1}{2} \partial_a \phi \partial^a \phi + \frac{ e^{\Phi}}{2\pi} C^{(1)}_a \partial^a \phi + \frac{1}{2} \left(  \frac{e^{\Phi}}{2\pi}  \right)^2 C^{(1)}_a C^{(1)a} \right\}.
\end{equation}
For the transverse fluctuations, it is convenient not only to use the new angular coordinates $\chi$, $\phi_{1}$ and $\phi_{2}$, but also
to write each of two 2-spheres in the $\mathbb{CP}^{3}$ in terms of the Euclidean coordinates
\begin{eqnarray}
  && u_{1} = \sin{\theta_{1}}\cos{\phi_{1}}, ~~~~~~
  u_{2} = \sin{\theta_{1}}\sin{\phi_{1}}, ~~~~~~
  u_{3} = \cos{\theta_{1}}, \nonumber\\
  && v_{1} = \sin{\theta_{2}}\cos{\phi_{2}}, ~~~~~~
  v_{2} = \sin{\theta_{2}}\sin{\phi_{2}}, ~~~~~~
  v_{3} = \cos{\theta_{2}}.\nonumber
\end{eqnarray}
Here $u_{1}^{2} + u_{2}^{2} + u_{3}^{2} = 1$ and $v_{1}^{2} + v_{2}^{2} + v_{3}^{2} = 1$, so that we can eliminate $u_{3}$ and $v_{3}$ in
favor of the other coordinates. In these coordinates, the $\textrm{AdS}_{4}\times\mathbb{CP}^{3}$ metric presents as
\begin{eqnarray}
  \nonumber & \!\!\! R^{-2}ds^{2} & = -\left(1+r^{2}\right)dt^{2} + \frac{dr^{2}}{(1+r^{2})}  
  + r^{2}\left(d\theta^{2}
  + \sin^{2}{\theta}d\varphi^{2}\right) + 4d\zeta^{2} \\
  \nonumber && + \cos^{2}{\zeta}\left[du_{1}^{2} + du_{2}^{2}
  + \frac{\left(u_{1}du_{1} + u_{2}du_{2}\right)^{2}}{1-\left(u_{1}^{2} +   
  u_{2}^{2}\right)}\right]
  + \sin^{2}{\zeta}\left[dv_{1}^{2} + dv_{2}^{2}
  + \frac{\left(v_{1}dv_{1} + v_{2}dv_{2}\right)^{2}}{1-\left(v_{1}^{2} + v_{2}^{2}\right)} 
  \right] \\
  \nonumber &&
  + 4\cos^{2}{\zeta}\sin^{2}{\zeta}\left[d\chi + \frac{1}{2}\left(\left[1-(u_{1}^{2} +   
  u_{2}^{2})\right]^{\tfrac{1}{2}}-1\right)
  \frac{\left(u_{1}du_{2}-u_{2}du_{1}\right)}{\left(u_{1}^{2} + u_{2}^{2}\right)} \right. \\
  && \left. ~~~~~~~~~~ ~~~~~~~~~~ ~~~~~~~~~~ 
  - \frac{1}{2}\left(\left[1-(v_{1}^{2} + v_{2}^{2})\right]^{\tfrac{1}{2}}-1\right)
  \frac{\left(v_{1}dv_{2}-v_{2}dv_{1}\right)}{\left(v_{1}^{2} + v_{2}^{2}\right)}\right]^{2}.
\end{eqnarray}
In what follows, we will again work in the gauge $\sigma_{0} \equiv \tau = t$, $\sigma_{1} = \theta$ and $\sigma_{2} = \varphi$, and write the fluctuations of the giant as
\begin{eqnarray}
  \nonumber && r = r_{0} + \varepsilon \delta r, ~~~~~~
  \chi = \omega_{0} t + \varepsilon \delta\chi, ~~~~~~
  \zeta = \tfrac{\pi}{4} + \varepsilon \delta\zeta, ~~~~~~~~ \nonumber\\
  && u_{i} = \varepsilon \delta u_{i}, ~~~~~~~~~~
  v_{i} = \varepsilon \delta v_{i}, ~~~~~~~~~~~~~ \phi = \varepsilon \delta \phi, \nonumber
\end{eqnarray}
with $i = 1,2$. To find the action for the fluctuations, we expand $S_{D2}$ to second order in $\varepsilon$.  Writing
$g = b_{0} + \varepsilon b_{1} + \varepsilon^{2} b_{2} + \dots$, we find that
\begin{eqnarray}
	S &\approx& 
	-\tfrac{1}{2} T_{2} k R^{2} \int d^3\sigma 
	\left\{ 
		\frac{\sqrt{-g}}{R^{3}}
		- \left( r_{0}^{3} + 3\varepsilon r_{0}^{2} \delta r + 3\varepsilon^{2} r_{0} \delta   
			r^{2}\right)\sin{\theta} 
	\right\} 
	+ S_\phi.
\end{eqnarray}
Here
\begin{equation}
  \frac{\sqrt{-g}}{R^{3}}
  \approx \sqrt{b_{0}} + \varepsilon \left(\tfrac{1}{2}\frac{b_{1}}{\sqrt{b_{0}}}\right)
  + \varepsilon^{2}\left(\tfrac{1}{2}\frac{b_{2}}{\sqrt{b_{0}}} - \frac{1}  
  {8}\frac{b_{1}^{2}}{b_{0}\sqrt{b_{0}}}\right),
\end{equation}
and we have defined
\begin{eqnarray}
  b_{0} &\equiv& \left(1+r_{0}^{2}-\omega_{0}^{2}\right)r_{0}^{4}\sin^{2}{\theta},   
  ~~~~~~~~~~ ~~~~~~~~~~ ~~~\nonumber\\
  b_{1} &\equiv& 2\left[\left(2+3r_{0}^{2}-2\omega_{0}^{2}\right)r_{0}^{3}\delta r -   
  \omega_{0}r_{0}^{4}\dot{\delta\chi}\right]\sin^{2}{\theta},\nonumber
\end{eqnarray}
\begin{eqnarray}
	\nonumber & b_{2} \equiv & 
		\frac{\sin^2 \theta}{(1 + r_{0}^{2})}
			\left[ -r_{0}^{4} \dot{\delta r}^{2} + 
			r_{0}^{2} \left( 1+r_{0}^{2} - \omega_{0}^{2} \right) (\nabla \delta r)^2 \right] \\
	\nonumber && + \sin^2 \theta 
		\left[ 9r_{0}^{4} + 6r_{0}^{2} \left( 1 + r_{0}^{2}-\omega_{0}^{2} \right) \right] \delta r^{2}
		-  \sin^{2}{\theta}  (8 \omega_{0} r_{0}^{3}) \dot{\delta\chi} \delta r  \\
	\nonumber && + \sin^2 \theta \left[ - r_{0}^{4} \dot{\delta\chi}^{2} +   
		r_{0}^{2} \left( 1 + r_{0}^{2} \right) \left( \nabla \delta\chi \right)^{2} 
		\right] \\
	\nonumber && + 4 \sin^{2}{\theta} \left[ - r_{0}^{4} \dot{\delta\zeta}^{2} +   
		\omega_{0}^{2} r_{0}^{4} \delta\zeta^{2}
		+  r_{0}^{2} \left( 1 + r_{0}^{2} - \omega_{0}^{2} \right) \left( \nabla \delta \zeta \right)^2 \right] \\
	\nonumber && + \frac{1}{2} \sin^2 \theta \sum_{i} 
		\left[ - r_{0}^{4} \dot{\delta u}_{i}^{2}  + r_{0}^{2} \left( 1 + r_{0}^{2} - \omega_{0}^{2} \right)
		\left( \nabla \delta u_i \right)^2 \right] \\
	\nonumber && + \frac{1}{2} \sin^2 \theta \sum_{i} 
		\left[ - r_{0}^{4} \dot{\delta v}_{i}^{2}  + r_{0}^{2} \left( 1 + r_{0}^{2} - \omega_{0}^{2} \right)
		\left( \nabla \delta v_i \right)^2 \right] \\
	&& + \frac{1}{2} \sin^{2} \theta  \omega_{0} r_{0}^{4} 
		\left(\delta u_{1}\dot{\delta u}_{2} - \delta u_{2} \dot{\delta u}_{1} \right)
	+ \frac{1}{2} \sin^{2}{\theta} \omega_{0} r_{0}^{4} 
		\left( \delta v_{1} \dot{\delta v}_{2} - \delta v_{2} \dot{\delta v}_{1} \right), \nonumber
\end{eqnarray}
and
\begin{equation}
	S_\phi = - \varepsilon^2 r_0 k T_{2}  \int d^3\sigma \sin \theta \left( \tfrac{4\pi^2 }{4} \left[ - \dot{\delta \phi}^2 + (\nabla \delta \phi)^2 \right] + (2\pi R \omega_0 ) \delta \zeta \dot{\delta \phi} - (R \omega_0 \delta \zeta)^2 \right)
\end{equation}
Organizing the expansion of the action in powers of $\varepsilon$
\begin{equation}
  S = S_{0} + \varepsilon S_{1} + \varepsilon^{2} S_{2} + \cdots.
\end{equation}
At zeroth order
\begin{equation}
  S_{0} = -a\int{dt}\left\{r_{0}^{2}\sqrt{1+r_{0}^{2}-\omega_{0}^{2}} 
  - r_{0}^{3}\right\},
\end{equation}
is of course just the D2-brane action evaluated on the dual giant configuration.
The first order action
\begin{equation}
  S_{1} = -\frac{a}{4\pi}\int dtd\theta d\varphi
  \left\{\left[\frac{(2+3r_{0}^{2}-2\omega_0^{2})r_{0}}{\sqrt{1+r_{0}^{2}-  
  \omega_{0}^{2}}} - 3r_{0}^{2}\right]\delta r
  -\frac{\omega_{0}r_{0}^{2}}{\sqrt{1+r_{0}^{2}-\omega_{0}^{2}}}~\dot{\delta\chi}\right  
  \}\sin{\theta}, ~~~~~~
\end{equation}
is less trivial. The second term in $S_{1}$ is easily recognised as a total derivative and can be dropped. For our zeroth order solution to satisfy the equations of motion, however,
the first term must vanish. This is a minimum when $r_{0} = 0$ (the point graviton) or $\omega_{0}^{2} = 1$ and $r_{0}^{2} = p^{2}$ (the giant graviton).

\noindent
After integrating by parts, the second order action is
\begin{eqnarray}
	\nonumber & S_{2} = & \frac{T_2 R^2 k r_0}{4} \int d^3\sigma \sin{\theta} \left\{
		\frac{\delta r}{(1+r_{0}^{2})} 
			\left[ - \ddot{\delta r} + \nabla^{2} \delta r \right]
		+ \frac{(1+r_{0}^{2})}{r_{0}^2} \delta \chi 
			\left[ -\ddot{\delta\chi} + \nabla^{2} \delta \chi \right] 
		+ 2 \dot{\delta\chi} \delta r  \right. \\
	\nonumber && ~~~~~~~~~~ ~~~~~~~~  
		+ \tfrac{1}{2} \sum_{i} \delta u_{i} \left[ - \ddot{\delta u_{i}} + \nabla^{2}\delta u_{i} \right]
  		+ \tfrac{1}{2} \left( \dot{\delta u_{1}} \delta u_{2} - \dot{\delta u_{2}} \delta u_{1} \right) \\
	&& ~~~~~~~~~~~~~~~~~~~
		+ \tfrac{1}{2} \sum_{i} \delta v_{i} \left[ - \ddot{\delta v_{i}} + \nabla^{2} \delta v_{i} \right] 
		+ \tfrac{1}{2} \left( \dot{\delta v_{1}} \delta v_{2} - \dot{\delta v_{2}} \delta v_{1} \right) \\
	\nonumber && ~~~~~~~~~~~ \left.
		+ 4 \delta\zeta \left[ - \ddot{\delta\zeta} + \nabla^{2} \delta\zeta \right] + 
		\left( \frac{2\pi}{R} \right)^2 \delta \phi \left[ - \ddot{\delta \phi} + \nabla^2 \delta \phi \right] -
		4 \left( \frac{2\pi}{R} \right) \delta \zeta \dot{\delta \phi} 
		\right\}
\end{eqnarray}
To proceed further, we decompose the perturbations into Fourier components
\begin{equation}
	\delta X (t, \theta, \varphi) = \delta X_{lm} (\omega) \ e^{-i\omega t} Y_{lm} (\theta, \phi)
\end{equation}
where $Y_{lm}(\theta,\phi)$ are spherical harmonics, satisfying
\begin{equation}
    \nabla^{2}Y_{lm}(\theta,\phi) = -Q_{l}Y_{lm}(\theta,\varphi), ~~~~~~ 
    \textrm{with} ~~ Q_{l} \equiv l(l+1).
\end{equation}
We vary the second order action to get the equations of motion, which decouple into 4 sets.
\begin{itemize}

\item \underline{Radial/Momentum modes}
\begin{equation*}
	\begin{pmatrix}
		\frac{r_0}{\left(1+ r_{0}^2\right)} \left( \omega^{2} - Q_{l} \right) & + i \omega \\
		- i \omega & \frac{\left(1+r_{0}^2\right)}{r_0} \left( \omega^{2} - Q_{l} \right)
	\end{pmatrix}
	\begin{pmatrix}
		\delta r \\
		\delta \chi \\
	\end{pmatrix} = 0
\end{equation*}
which has non-zero solutions only when the determinant of the matrix vanishes.  Consequently, the normal modes oscillate with frequencies
\begin{equation}
	\omega^{(r)}_{\pm} = 
  	\begin{pmatrix}
		l \\
		l+1
	\end{pmatrix}.
\end{equation}
\item \underline{$S^2 \times S^2$ modes}
\begin{equation}
	\begin{pmatrix}
    		\omega^{2} - Q_{l} & - i \omega \\
    		+ i \omega &  \omega^{2} - Q_{l}
	\end{pmatrix}
	\begin{pmatrix}
		\delta u_{1} \\
    		\delta u_{2} \\
  	\end{pmatrix} =
	\begin{pmatrix}
    		\omega^{2} - Q_{l} & - i \omega \\
    		+ i \omega &  \omega^{2} - Q_{l}
  	\end{pmatrix}
  	\begin{pmatrix}
    		\delta v_{1} \\
    		\delta v_{2} \\
  	\end{pmatrix} = 0,\nonumber
\end{equation}
leading to a set of normal modes for each two-sphere - corresponding to $u(i=1)$ and $v(i=2)$,
\begin{equation}
  \omega_\pm^{(i)} =\begin{pmatrix}
		l \\
		l+1
	\end{pmatrix}
	\hspace{3cm}
	i=1,2.
\end{equation}
\item \underline{$S^1$ radius/gauge field}
\begin{equation*}
	\begin{pmatrix}
		2 \left( \omega^{2} - Q_{l} \right) & - i \omega \\
		+ i \omega & \tfrac{1}{2} \left( \omega^{2} - Q_{l} \right)
	\end{pmatrix}
	\begin{pmatrix}
		R\delta \zeta \\
		2\pi \delta \phi \\
	\end{pmatrix} = 0
\end{equation*}
so that the normal modes have frequencies
\begin{equation}
	\omega^{(r)}_{\pm} = 
  	\begin{pmatrix}
		l \\
		l+1
	\end{pmatrix}.
\end{equation}
\end{itemize}
In all cases, the spectrum of fluctuations is entirely real (there are no tachyonic modes), from which we conclude that, like its D3-brane counterpart in $AdS_{5} \times S^{5}$, the dual giant graviton on $AdS_{4} \times \mathbb{CP}^{3}$ is perturbatively stable. At this point, it is worth noting that:
\begin{itemize}
\item
Any dependence of the spectrum on the size of the giant would be a
trace signature of the geometry that could be probed in the dual field theory. 
However, we see here that frequencies of fluctuations are independent of the radius of the giant.   Physically, this can be attributed to the exact cancellation of two  
competing effects: (1) the blueshifting of frequency from rising within the AdS gravitational well and 
(2) the increased wavelength on a larger worldvolume\footnote{We thank R. de Mello Koch for pointing 
this argument out to us.} \cite{Giant-results}.
\item
The spectrum contains massless goldstone modes from the breaking of a number of continuous symmetries. The radial part of the $SO(2,3)$ AdS symmetry is broken by the  choice of radius/momentum of the giant, leading to a goldstone mode on the worldvolume with frequency $\omega^{(r)}_-=l$. There is also a broken $SU(2) \times SU(2) \subset SU(4)$ symmetry, corresponding to the $S^2 \times S^2 \subset \mathbb{CP}^3$ - the two normal modes with frequencies $\omega^{(i)}_-=l$ are the corresponding goldstone bosons. These broken symmetries (and the resulting massless modes) can be easily seen in the gauge theory as well.
\item
Perhaps most intriguingly, there is also a massless mode coming from the coupling of the gauge field and the radius of the $S^1$ direction of motion, $\zeta$. This implies that there are giant graviton solutions (with the same energy) that involve non-trivial gauge fields. Such solutions can also be thought of as D0-brane charge dissolved in the giant worldvolume. In fact, using the Poincar\'e dual of the scalar field, it is not hard to see that the infinitesimal worldvolume gauge flux is
\begin{eqnarray}
	F &=& - *( d\phi + \frac{e^{\Phi}}{2\pi} C^{(1)})  \\  
		 &=& - \frac{ 2R \varepsilon \ \delta \zeta }{2\pi} \left( Rr_0 \sin \theta d\theta \wedge d\varphi \right)  \nonumber
\end{eqnarray}
which is the flux for a Dirac monopole of charge $(-2R \varepsilon \delta \zeta ) / 2\pi$.  Interestingly, such solutions would seem to have a maximum D0-charge, found when the radius of the $S^1$ direction of motion shrinks to zero size.
Similar solutions were found in \cite{Takayanagi-rings}, though without $\mathbb{CP}^3$-momentum charge.  It would be interesting to find the relevant charged dual D2-brane giants in this case.
\end{itemize}

%-------------------------------------------------------------------------------------
\section{Open strings attached to the dual giant graviton}
\label{Open strings}
%-------------------------------------------------------------------------------------
Transverse fluctuations of D-branes are coded in open strings attached to and moving on the brane. In this section, we will explicitly study such strings from the string worldsheet perspective. As emphasized in the introduction to this article, a full quantum treatment of the worldsheet sigma model is sorely lacking. However several interesting and instructive limits exist. Two in particular will be of interest to us; short strings and long strings. In this section, we extend the results of \cite{Berenstein-et-al, Correa-silva} and present a systematic treatment of both these limits to the D2-brane giant graviton on $AdS_{4} \times \mathbb{CP}^{3}$.
To study short strings on the dual giant, we need to take a limit in which the size of the string is ``amplified" with respect to the geometry that it probes. In 
\cite{Berenstein-et-al} it was argued that a good description of these short strings (attached to a non-maximal sphere giant) is of open strings attached to a flat D3-brane in a pp-wave background. This was later extended to open strings on AdS giants in \cite{Correa-silva} with qualitatively similar results. The pp-wave limit is particularly useful since there the string action is quadratic in lightcone-gauge and consequently solvable. In what follows, we quantize the short string sigma model on the pp-wave associated to the D2-brane giant and compute its bosonic spectrum exactly before moving on to long semiclassical strings.\\
%----------------------------------------------------------------------------------
\subsection{Short strings in a pp-wave geometry} \label{short strings}
%----------------------------------------------------------------------------------
\noindent
To take the Penrose limit, it is most convenient to work in global AdS coordinates in which the radial coordinate $r = \sinh{\rho}$, and the AdS$_{4}\times\mathbb{CP}^{3}$ metric reads\begin{eqnarray}
  \nonumber & R^{-2} ds^{2} = & -\cosh^{2}{\rho}\,dt^{2} + d\rho^{2}
  + \sinh^{2}{\rho}\left(d\theta^{2} + \sin^{2}{\theta}\,d\varphi^{2}\right) \\
  \nonumber && + 4d\zeta^{2} + \cos^{2}{\zeta}\left(d\theta_{1}^{2}+\sin^{2}{\theta_{1}}  
  \,d\phi_{1}^{2}\right)
  + \sin^{2}{\zeta}\left(d\theta_{2}^{2}+\sin^{2}{\theta_{2}}\,d\phi_{2}^{2}\right) \\
  && + 4\sin^{2}{\zeta}\cos^{2}{\zeta}\left[d\chi - \sin^2{\frac{\theta_{1}}{2}}\,d\phi_{1}
  + \sin^2{\frac{\theta_{2}}{2}}\,d\phi_{2} \right]^{2}. ~~~~~~
\end{eqnarray}
The null geodesic defined by
\begin{equation}
  t = \chi = \varphi = u, ~~~~ \theta = \tfrac{\pi}{2}, ~~~~ \rho = \rho_{0}, ~~~~ \zeta =   
  \tfrac{\pi}{4} ~~~ \textrm{and}
  ~~~ \theta_{i} = 0,\nonumber
\end{equation}
describes a trajectory parameterized by $\varphi = u$ on the dual giant graviton.  To construct the pp-wave geometry associated with this null
geodesic, we take the ansatz
\begin{eqnarray} \label{pp ansatz1}
  \nonumber && t = u + \frac{v}{R^{2}\cosh^{2}{\rho_{0}}},\\
   && \chi = u - \frac{v}{R^{2}\cosh^{2}{\rho_{0}}} -   
  \frac{\tanh{\rho_{0}}~y_{1}}{R}, \\
  \nonumber && \varphi = u - \frac{v}{R^{2}\cosh^{2}{\rho_{0}}} + \frac{y_{1}}{R\cosh{\rho_{0}}  
  \sinh{\rho_{0}}},
\end{eqnarray}
and expand about the geodesic as
\begin{eqnarray} \label{pp ansatz2}
  && \rho = \rho_{0} + \frac{y_{2}}{R}, ~~~~~~
  \theta = \frac{\pi}{2} + \frac{z_{1}}{R\sinh{\rho_{0}}}, ~~~~~~
  \zeta = \frac{\pi}{4} + \frac{z_{2}}{2R}, ~~~~~~
  \theta_{i} = \frac{\sqrt{2}r_{i}}{R}, ~~~~~~
\end{eqnarray}
with $\phi_{i}$ unspecified. We can now take the Penrose limit, in which $R$ is large and 
we zoom in on the null geodesic.  
With $r_{0} \equiv \sinh{\rho_{0}}$ fixed, the radius of the dual giant $\sim R$ and consequently diverges. As in the $AdS_{5} \times S^{5}$ case, this short string limit is effectively just a treatment of open strings propagating on a flat D2-brane on the pp-wave
\begin{eqnarray}
  \nonumber & ds^{2} = & -4dudv - \left(\sum_{i=1}^{2}z_{i}^{2}\right)du^{2} +   
  \sum_{i=1}^{4}{dx_{i}^{2}} + \sum_{i=1}^{2}{dy_{i}^{2}} + \sum_{i=1}^{2}{dz_{i}^{2}}   
  \\&& + \left[ 4y_{2}dy_{1}  + x_{2}dx_{1} - x_{1}dx_{2} + x_{4}dx_{3} - x_{3}dx_{4}   
  \right] du,
\end{eqnarray}
where, for convenience, we have chosen to write it in terms of Cartesian coordinates $x_{2k-1} = r_{k}\,\cos\phi_{k}$ and $x_{2k} = r_{k}\,\sin\phi_{k}$ (with $k=1,2$). In this form, the similarity that this pp-wave bears to the homogeneous plane wave of \cite{Homogeneous pp-wave} is manifest\footnote{See also the second of \cite{Fluctuations} for further discussion of the relation between the homogeneous plane wave and the standard pp-wave in magnetic coordinates.} and the techniques developed there are easily adapted to this case. In lightcone gauge $u=2p^{u}\tau$, the bosonic part of the Polyakov action is
\begin{eqnarray}
  \nonumber & \!\!\! S = & \int{d\tau}\int_{0}^{\pi}{\frac{d\sigma}{2\pi}}
  \left\{\sum_{I=1}^{4}\left(\tfrac{1}{2}\dot{X}_{I}^{2} - \tfrac{1}{2}X_{I}'^{2}\right)
  + \tfrac{1}{2}m \left(X_{2}\dot{X}_{1} - X_{1}\dot{X}_{2}\right) + \tfrac{1}{2}m   
  \left(X_{4}\dot{X}_{3} - X_{3}\dot{X}_{4}\right) \right. \\
  && \left. ~~~~~~~~~~ ~~~~~~ + \sum_{I=1}^{2}\left(\tfrac{1}{2}\dot{Y}_{I}^{2} -   
  \tfrac{1}{2}Y_{I}'^{2}\right) + 2m Y_{2}\dot{Y}_{1}
  + \sum_{I=1}^{2}\left(\tfrac{1}{2}\dot{Z}_{I}^{2} - \tfrac{1}{2}Z_{I}'^{2} + \tfrac{1}  
  {2}m^{2}Z_{I}^{2}\right) \right\}, ~~~~~~~~
\end{eqnarray}
with $m \equiv 2p^{u}$.  Note that the string embedding coordinates $X_{I}$, associated with the two 2-spheres, are coupled in pairs,
\begin{equation}
	X_{\pm}^{(i)} = X_{2i-1} \pm iX_{2i},
\end{equation}
as are the $Y_I$, which descend from $\rho$ and $\chi$
\begin{equation}
	Y_{\pm} = Y_{1} \pm iY_{2}.
\end{equation}
In the large $R$ limit, the open string boundary conditions associated with the AdS giant imply Dirichlet boundary conditions on
$X_{I}$, $Y_{I}$ and $Z_{2}$, and Neumann boundary conditions\footnote{Note that the lightcone gauge
choice is consistent with Neumann boundary conditions on $U$.} on the lightcone coordinates $U$ and $V$, as well as $Z_{1}$.
Solving the open string equations of motion - subject to the appropriate boundary conditions - and quantizing the bosonic sector, we obtain the following expressions
for the Neumann embedding coordinate
\begin{eqnarray}
Z_{1}(\tau,\sigma) &=& \sqrt{\frac{1}{m}} 
	\left[ \xi_{0}^{1}e^{-im\tau} +  \left( \xi_{0}^{1} \right)^{\dag} e^{im\tau} \right]
	+ \sum_{n=1}^{\infty} \sqrt{\frac{2}{\omega_{n}}}
	\left[
		\xi_{n}^{1} e^{-i\omega_{n}\tau} +  \left(\xi_{n}^{1}\right)^{\dag} e^{i\omega_{n}\tau}
	\right] \cos(n\sigma) \nonumber
\end{eqnarray}
and the Dirichlet coordinates
\begin{eqnarray}
X_+^{(i)}(\tau,\sigma) &=&  \sum_{n=1}^{\infty}\sqrt{\frac{4}{\tilde{\omega}_n}}
	  \left[
	  	\alpha_{n}^{(i-)} e^{- i\tilde{\omega}^{-}_{n}\tau} +
	  	\left( \alpha_{n}^{(i+)}\right)^{\dagger} e^{ i\tilde{\omega}^{+}_{n}\tau}
	\right] \sin(n\sigma)
	\nonumber \\
X_{-}^{(i)}(\tau,\sigma) &=&  \sum_{n=1}^{\infty}\sqrt{\frac{4}{\tilde{\omega}_n}}
	  \left[
	  	\alpha_{n}^{(i+)} e^{- i\tilde{\omega}^{+}_{n}\tau} +
	  	\left( \alpha_{n}^{(i-)} \right)^{\dagger} e^{ i\tilde{\omega}^{-}_{n}\tau}
	\right] \sin(n\sigma)
	\nonumber \\
Y_+(\tau,\sigma) &=&  \sum_{n=1}^{\infty}\sqrt{\frac{4}{\omega_n}}
	  \left[
	  	\beta_{n}^{-} e^{- i\omega^{-}_{n}\tau} +
	  	\left( \beta_{n}^{+} \right)^{\dagger} e^{ i\omega^{+}_{n}\tau}
	\right] \sin(n\sigma) \\
Y_{-}(\tau,\sigma) &=&  \sum_{n=1}^{\infty}\sqrt{\frac{4}{\omega_n}}
	  \left[
	  	\beta_{n}^{+} e^{- i\omega^{+}_{n}\tau} +
	  	\left( \beta_{n}^{-} \right)^{\dagger} e^{ i\omega^{-}_{n}\tau}
	\right] \sin(n\sigma)
	\nonumber \\
Z_{2}(\tau,\sigma) &=& \sum_{n=1}^{\infty} \sqrt{\frac{2}{\omega_{n}}}
	\left[
		\xi_{n}^{2} e^{-i\omega_{n}\tau} + \left(\xi_{n}^{2}\right)^{\dag} e^{i\omega_{n}\tau}  
	\right] \sin(n\sigma). \nonumber
\end{eqnarray}
Here,
\begin{eqnarray}
\nonumber && \omega_{n} \equiv \sqrt{m^{2}+n^{2}}
			~~~~~~~~~~
			\tilde{\omega}_{n} \equiv \sqrt{\tfrac{1}{4} m^{2} + n^{2}} \\
\nonumber && \omega_{n}^{\pm} \equiv \omega_{n} \pm m
			~~~~~~~~~~~~~
			\tilde{\omega}_{n}^{\pm} \equiv \tilde{\omega}_{n} \pm \tfrac{1}{2}m
\end{eqnarray}
are the string frequencies,
and the creation and annihilation operators satisfy the usual commutation relations
\begin{equation}
	[ \alpha_{n}^{i+}, \left( \alpha_{k}^{j+} \right)^{\dag} ] = 
	[ \alpha_{n}^{i-}, \left( \alpha_{k}^{j-} \right)^{\dag} ]  = 
	[ \xi_{n}^{i}, \left( \xi_{k}^{j} \right)^{\dag} ] =
	\delta^{ij} \delta_{nk} \hspace{1cm} i=1,2 \nonumber
\end{equation}
\begin{equation}
	[ \beta_{n}^{+}, \left( \beta_{k}^{+} \right)^{\dag} ] = 
	[ \beta_{n}^{-}, \left( \beta_{k}^{-} \right)^{\dag} ] = \delta_{nk}
\end{equation}
To determine the string spectrum, we express the lightcone Hamiltonian 
\begin{eqnarray}
\nonumber & \!\!\! H_{lc} = & \frac{1}{m}\int_{0}^{\pi}{\frac{d\sigma}{2\pi}}
\left\{\sum_{I=1}^{4}\left(\tfrac{1}{2}\dot{X}_{I}^{2} + \tfrac{1}{2}X_{I}'^{2}\right)
+ \sum_{I=1}^{2}\left(\tfrac{1}{2}\dot{Y}_{I}^{2} + \tfrac{1}{2}Y_{I}'^{2}\right)
+ \sum_{I=1}^{2}\left(\tfrac{1}{2}\dot{Z}_{I}^{2} + \tfrac{1}{2}Z_{I}'^{2} + \tfrac{1}{2}m^{2}Z_{I}^{2}\right)\right\},
\end{eqnarray}
in the normal ordered harmonic oscillator basis as
\begin{equation}
	H_{lc} = 
	 \left( \xi_{0}^{1} \right)^{\dag} \xi_{0}^{1} +
	 \sum_{n=1}^{\infty}
	\left[
		\frac{\omega_{n}}{m}
		\left(
			\left( \xi_{n}^{1} \right)^{\dag} \xi_{n}^{1} +
			\left( \xi_{n}^{2} \right)^{\dag} \xi_{n}^{2}
		\right) +
		\left(
			\frac{\omega_{n}^{+}}{m} \left( \beta_{n}^{+} \right)^{\dag} \beta_{n}^{+} + 
			\frac{\omega_{n}^{-}}{m} \left( \beta_{n}^{-} \right)^{\dag}  \beta_{n}^{-}
		\right) +
	\right. \nonumber
\end{equation}
\begin{equation}	
	\left.
	\sum_{i=1,2}
		\left(
			\frac{ \tilde{\omega}_{n}^{+}}{m} \left( \alpha_{n}^{i+} \right)^{\dag} \alpha_{n}^{i+} +
			\frac{ \tilde{\omega}_{n}^{-}}{m} \left( \alpha_{n}^{i-} \right)^{\dag} \alpha_{n}^{i-}
		\right)	
	\right] .
\end{equation}
Let us now relate the mass, $m$, of these strings to parameters in the original AdS space. 
The relevant AdS quantities are the energy and momenta of the string ($E, J_\chi$, and $J_\varphi$). In the Penrose limit, these translate to 
the lightcone momenta
\begin{equation}
H_{lc} = -p_{u} = E - (J_{\chi} + J_{\varphi}) ~~~~~ \textrm{and} ~~~~~
-p_{v} = \frac{E+(J_{\chi}+J_{\varphi})}{R^{2}\cosh^{2}{\rho_{0}}} = m ~~
\end{equation}
and the spatial momenta in the $y_1$ direction
\begin{equation}
p_{y_{1}} = -\frac{1}{R}\tanh{\rho_{0}}\left(J_{\chi} - \frac{1}{\sinh^{2}{\rho_{0}}}J_{\varphi}\right).
\end{equation}
Keeping $p_{y_1}$ finite, we require $J_{\chi} = \tfrac{1}{\sinh^{2}{\rho_{0}}} J_{\varphi} + \mathcal{O}(R)$.  To keep the lightcone Hamiltonian finite, $E = J_{\chi} + J_{\varphi} + \mathcal{O}(1)$, so that
\begin{equation}
J_{\varphi} \equiv L, ~~~~~~ J_{\chi} = \frac{L}{\sinh^{2} \rho_{0} } (1 + \mathcal{O}(R/L))~~~~ \textrm{and} ~~~~ E = L \coth^{2}\rho_{0}\, (1 + \mathcal{O}(R/L) ).
\end{equation}
This gives an inverse mass to the string excitations
\begin{eqnarray}
	\frac{1}{m^{2}} &=& \frac{1}{p_v^2} \\
    			&=&\frac{R^4 \cosh^4 \rho_0}{(E+ J_\chi + J_\varphi)^2} \nonumber\\
    			& = &  \frac{\pi^{2} \lambda}{2L^{2}} \sinh^{4}{\rho_{0}} \left( 1 + \mathcal{O} (  R/L ) \right)
			\nonumber
\end{eqnarray}
in terms of the 't Hooft coupling $\lambda = \frac{R^{4}}{2\pi^{2}}$ \cite{ABJM}.  In the limit in
which $\tilde{\lambda} \equiv \tfrac{\lambda}{L^{2}}$ is fixed and small, the energy eigenvalues of the lightcone Hamiltonian can be
approximated as:
\begin{eqnarray}
& \alpha_{n}^{i \pm}: ~~~~ 
	& \frac{\tilde{\omega}_{n}^{\pm}}{m} = 
		\sqrt{\tfrac{1}{4} + \tfrac{n^{2}}{m^{2}}} \pm \tfrac{1}{2}
		\approx 1 + \frac{n^{2}}{m^{2}} ~~ \textrm{or} ~~ \frac{n^{2}}{m^{2}}, ~~~~~ \nonumber \\
& \beta_{n}^{\pm}: ~~~~ 
	& \frac{\omega_{n}^{\pm}}{m} = 
		\sqrt{1 + \tfrac{n^{2}}{m^{2}}} \pm 1
		\approx 2 + \frac{n^{2}}{2m^{2}} ~~ \textrm{or} ~~ \frac{n^{2}}{2m^{2}}, ~~~~~ \\
& \xi_{n}^{i}: ~~~~ 
	& \frac{\omega_{n}}{m} = 
		\sqrt{1 + \tfrac{n^{2}}{m^{2}}}
		\approx 1 + \frac{n^{2}}{2m^{2}}, ~~~~~ \nonumber 
\end{eqnarray}
which can clearly be organized in powers of $\tilde{\lambda}$, including states that are nearly massless (in the limit $\tilde{\lambda} \to 0$).\\ 

\noindent
In the canonical AdS${}_5 \times S^5$ background, the existence of the finite $\tilde{\lambda} = \lambda/L^2$ scaling limit is tied to the BMN scaling in the gauge theory \cite{BMN}.
In this limit, it is possible to find ``near-BPS'' states and compute their anomalous dimensions, which match (at least to third order) with the exact string modes.  A comparison for open string modes was done in \cite{Berenstein-et-al}. However, the pp-wave limit of closed strings has been compared to the BMN limit of the Chern-Simons gauge theory in \cite{pp-wave}, and they were found not to match for finite $\tilde{\lambda}$.  It seems that, due to the lower amount of symmetry in this background, BMN scaling is violated, and the anomalous dimension of BMN operators should scale as
\begin{equation*}
	\delta_n = f(\lambda) \frac{n^2}{L^2}
\end{equation*}
where the function $f(\lambda)$ has different limits with $\lambda$
\begin{eqnarray*}
	f(\lambda) &\sim \lambda & \hspace{1cm} \lambda \to \infty  \\
	f(\lambda) &\sim \lambda^2 & \hspace{1cm} \lambda \to 0
\end{eqnarray*}
It would be interesting to see the analogous scaling in the case of open strings.

%-------------------------------------------------------------------------------------
\subsection{Long semiclassical strings} \label{long strings}
%-------------------------------------------------------------------------------------

Zooming back out to the full AdS space, we can also consider {\it long strings} ending on 
the dual giant graviton. Even though the string worldsheet on $AdS_{4} \times \mathbb{CP}^{3}$ is every bit as difficult to quantize as the usual $AdS_{5} \times S^{5}$ case, we can still look at the worldsheet action of a subspace of these string states, which facilitate comparison with a semiclassical analysis of the spin chain.  For strings moving along the worldvolume of the D2-brane, the analysis (at least on the gravity side) is identical to that carried out for dual D3-brane giants in $AdS_{5}$, so we will restrict ourselves to a discussion of the final result of our computation and any differences between this case and the D3-brane and refer the interested reader to \cite{Correa-silva} for details.

\noindent
The idea is that we restrict to a string propagating on an $AdS_{3} \times S^{1} \subset AdS_{4} \times \mathbb{CP}^3$ at $\xi=\frac{\pi}{4}$, so that
\begin{equation}
	R^{-2} ds^2 = -(1+r^2) \ dt^2 + \frac{dr^2}{1+r^2} + r^2 \ d\varphi^2 + d\chi^2 \;.
\end{equation}
The D2-brane is located at $r=p$ (the giant's angular momentum) and $\chi = t$ (the $S^1$ coordinate), leading to appropriate Dirichlet boundary conditions.  The key to simplifying the action lies in choosing the correct worldsheet gauge \cite{Kruczenski-et-al}.  Keeping in mind an eventual comparison to the gauge theory, we would like to work in the static gauge, $t=\tau$, so that the worldsheet and spacetime energies coincide (being dual to the anomalous dimension of gauge theory operators).  We also choose a gauge, $p_\varphi = 2 L$, in which the worldsheet momentum along the trajectory is constant along the string. Here, $L=\frac{1}{2\pi} \int_0^\pi \ d\sigma \ p_\varphi$ is dual to the spin of the gauge theory operator.  This choice of gauge has a subtle interpretation in the canonical $AdS_{5} \times S^5$ case - in deriving the spin chain from gauge theory operators, a choice must be made regarding the separation of the operator into ``sites'' of the spin chain, and choosing to spread the AdS  momentum density evenly along the worldsheet leads to a spin chain where the sites are organized according to worldvolume spin (i.e., each covariant derivative in the operator corresponds to a site in the spin chain).  We expect similar gauge effects here, when a systematic study of the dual gauge theory operators is completed.

\noindent
In the ``fast string'' limit, in which we focus on the trajectory $\phi=\chi-t = 0$, we assume that $\partial_\tau x^\mu \sim \lambda/L^2 \ll 1$, and the action reduces to
\begin{eqnarray}
 S = -L \int \ d\tau d\sigma\,\left[\frac{\dot{\phi}}{r^2-1} - 
 \frac{\lambda}{4L^2} (r'^2 + r^2 \phi'^2) + \mathcal{O} (\lambda^{2}/L^{4})\right] \;.
 \label{Polyakov}
\end{eqnarray}
In the AdS${}_5 \times S^5$ background, this action has been exactly matched to the semiclassical, Landau-Lifshitz action derived in the coherent state basis of the dual $\mathfrak{sl}(2)$ spin chain \cite{Correa-silva}, and we would expect that a similar matching should appear for the analogous limit in our case, modulo one very important caveat\footnote{We would like to thank the anonymous referee for bringing this point to our attention.}; the $\mathfrak{sl}(2)$ sector of the $\mathcal{N}=6$ Chern-Simons-matter theory, unlike $\mathcal{N}=4$ SYM {\it is not closed}. Indeed, at the level of the closed string $OSp(2,2|6)$ spin chain found in \cite{Minahan-Zarembo1}, it was reported that operators of the form $D_{\mu}Y_{A}^{\dagger}Y^{B}$ mix with fermionic operators $\bar{\psi}^{B}\gamma_{\mu}\psi_{A}$. One interpretation of this mixing is that covariant derivative excitations do not correspond to elementary magnons on the closed string spin chain but should instead be thought of as bound states of fermionic magnons.  It was shown that only in the \emph{strictly infinite} strong coupling limit do these excitations look independent \cite{Minahan-Zarembo2}, so it is expected that when string corrections are accounted for, they will dissolve into fermions.  Understanding this phenomenon in the open string sector, and extending the semiclassical action (\ref{Polyakov}) to include string corrections, is certainly an interesting and important pursuit but beyond the scope of this article.

%------------------------------------------------------------------------------------
\section{Discussion}
%------------------------------------------------------------------------------------
This article catalogues some features of the giant graviton phenomenon in the $AdS_{4}\times \mathbb{CP}^{3}$. In particular, we have focused on the ``dual" D2-brane giant gravitons blown up on an $S^{2}\subset AdS_{4}$ and their excitations. We have found that, just as for the dual D3-brane giant, all the fluctuation modes have real, positive $\omega^{2}$ that are manifestly independent of the size of the giants. This absence of any tachyonic modes in the spectrum again means that there are no perturbative instabilities in the configuration.  A particularly interesting aspect of the fluctuation spectra is the existence of coupling between the worldvolume gauge field and transverse fluctuations of the brane - this is a significant difference from the canonical dual D3-branes in AdS${}_5$. Similar couplings have been found in, e.g., \cite{Pirrone}, but here the coupling results in a massless goldstone mode, indicating a new type of dual D2-brane which has both momentum and D0-brane charge.

\noindent
Motivated by the remarkable insights yielded by similar studies of giant gravitons in $AdS_{5} \times S^{5}$ \cite{deMelloKoch1, Giant-results, Berenstein-et-al, Correa-silva}, in section 5 we presented an analysis of open strings attached to the D2-brane giant in the limit of long and short strings. In the latter, we confirmed that the families of null geodesics contained in the giant graviton trajectories found in \cite{Berenstein-et-al, Correa-silva} persist at least for the dual giants in $AdS_{4}\times \mathbb{CP}^{3}$. Consequently, we were able to take a standard Penrose limit about this geodesic and quantize the short open strings on the resulting pp-wave. Our findings here are in excellent agreement with what was reported for D3-brane giants; namely that here too the spectral structure of these open string excitations have a perturbative expansion in $\lambda/L^2$ for angular momentum $L\to \infty$. In the gauge theory, this suggests a potential simplification in the `BMN' or thermodynamic limit for the open spin chain. However, evidence from the closed string side implies a breaking of `BMN scaling' at one-loop. To study {\it long} open strings attached to the giant, we took a semiclassical limit in coordinates adapted to the D-brane and, in the fast string limit, computed to leading order in $\lambda/L^{2}$, the Polyakov action in (\ref{Polyakov}). To this order at least, we found complete agreement with the more familiar D3-brane giants.\\ 

\noindent
Although these similarities between the dual D3-brane giant and its D2-brane counterpart on $AdS_{4} \times \mathbb{CP}^{3}$ might seem mundane, it is worth keeping in mind two important points: 
\begin{itemize}
\item
The three-dimensional $\mathcal{N}=6$ super Chern-Simons theory conjectured to be 
dual to the type IIA string on $AdS_{4} \times \mathbb{CP}^{3}$ is decidedly different   
from $\mathcal{N}=4$ super Yang-Mills in field content, number of supersymmetries and 
even degrees of freedom. So, {\it a priori}, it is certainly not clear that agreement between
these two systems should be anything more than qualitative. Yet, the agreement we find is
quite precise.
\item
The D2-brane studied in this note is a descendant of an M2-brane in the IIA reduction. It would seem then, that the dynamics of the dual giant graviton on 
$AdS_{4}$ encodes in a nontrivial way, {\it membrane} dynamics in M-theory. 
In this context, the similarities between this giant graviton and the dual giant on $AdS_{5}$ that we outline here are all the more remarkable.
\end{itemize}
Clearly, there remains much to be learnt in the open string sector of the AdS/Chern-Simons duality. For example; perhaps the most concrete extension to this work is the matching of our string results with the dual $\mathcal{N}=6$ Chern-Simons theory.  Open strings (which includes the point-like fluctuations of section 3) are dual to `words' made of strings of gauge fields which have uncontracted indices.  To make them gauge invariant, these open strings must be attached to D-brane operators - in the case of dual giant gravitons, these are given by totally symmetric polynomials in the bi-fundamental scalar fields \cite{Schur}. The anomalous dimensions of such open string operators can be interpreted as the Hamiltonian for a spin chain with boundaries, and compared with gravity results \cite{deMelloKoch1, Correa-silva}.  A detailed study of open string operators in the AdS${}_4$/CFT${}_3$ correspondence  - along similar lines as \cite{Giant-results,More-giant-results} - would illuminate not only more of the gauge/gravity duality but perhaps also the structure of M-theory itself. 

%------------------------------------------------------------------------------------
\section{Acknowledgements}
%------------------------------------------------------------------------------------
We are most grateful to David Berenstein, Diego Correa, Sean Hartnoll, C. Krishnan, Raphael Nepomechie, Kostas Skenderis, and especially Robert de Mello Koch for useful discussions. J.M. acknowledges support from the NRF Thuthuka program under grant GUN61699 as well as an NRF Key International Scientific Collaboration grant under which much of this work was done. A.H. is supported by a Vice-Chancellor's Postdoctoral Fellowship of the University of Cape Town, A.P. by an NRF Scarce Skills PhD fellowship and M.S. by a National Institute for Theoretical Physics honours fellowship.

\end{document}